# The Discrete Noise of Magnons

**S. Rumyantsev[1,2,3], M. Balinskiy[1,2], F. Kargar[1,2], A. Khitun[1,2] and A. A. Balandin[1,2,*]**

[1]Department of Electrical and Computer Engineering and Phonon Optimized Engineered Materials (POEM) Center, University of California, Riverside, California 92521 USA

[2]Spins and Heat in Nanoscale Electronic Systems (SHINES) Center, University of California, Riverside, California 92521 USA

[3]Ioffe Physical-Technical Institute, St. Petersburg 194021 Russia

**Magnonics is a rapidly developing subfield of spintronics, which deals with devices and circuits that utilize spin currents carried by magnons – quanta of spin waves[1–7]. Magnon current, _i.e._ spin waves, can be used for information processing, sensing, and other applications. A possibility of using the amplitude and phase of magnons for sending signals via _electrical insulators_ creates conditions for avoiding Ohmic losses, and achieving ultra-low power dissipation[2–10]. Most of the envisioned magnonic logic devices are based on spin wave interference, where the minimum energy per operation is limited by the noise level[8,11] The sensitivity and selectivity of magnonic sensors is also limited by the low frequency noise[9,10]. However, the fundamental question "do magnons make noise?" has not been answered yet. It is not known how noisy magnonic devices are compared to their electronic counterparts. Here we show that the low-frequency noise of magnonic devices is dominated by the random telegraph signal noise rather than $1/f$ noise – a striking contrast to electronic devices ($f$ is a frequency). We found that the noise level of surface magnons depends strongly on the power level, increasing sharply at the on-set of nonlinear dissipation. The presence of the random telegraph signal noise indicates that the current fluctuations involve random discrete _macro_ events. We anticipate that our results will help in developing the next generation of magnonic devices for information processing and sensing.**

---

[*] Corresponding author (AAB): balandin@ece.ucr.edu

A possibility of using magnon currents in electrical insulators for information processing and various sensing applications generated excitement across many disciplines[2–10]. The most attractive feature is a prospect of avoiding Ohmic losses and associated Joule heating in electrically insulating magnetic materials[2]. Numerous devices with spin waves, *i.e.* magnon currents, have been experimentally demonstrated and compared to their electrical counterparts[1-14]. While thermal dissipation is an important characteristic of any device technology, there is another crucial metric, which has not been properly addressed in magnonic devices yet. We still do not know *how much noise magnon currents make* and *how different the noise of magnons from that of electrons*. These intriguing questions are interesting from both fundamental science and practical applications points of view. Only recently, theoretical studies on the specific noise types of spin currents started to appear[15]. However, no experimental investigations of the noise of magnons in electrically insulating spin waveguides have been reported. An urgent need to explore this important characteristic for magnonic devices motivates the present study.

The noise in electronic devices, made from metals and semiconductors, can be viewed as various manifestations of the *discreteness* of charges[16,17]. The Johnson – Nyquist thermal noise is associated the random thermal agitation of electrons while the shot noise is related to random events of electrons going over a potential barrier[16,17]. The low-frequency $1/f$ noise and generation-recombination (G-R) noise in semiconductors are related to the random process of individual electron caption and emission by the traps associated with defects (*f* is the frequency)[16–18]. With the electron Fermi wavelength $\lambda_F = 2\pi/k_F = 2\pi/(3\pi^2 n)^{1/3} \sim 0.1 - 0.5\ nm$, the notion of electrons as particles work well for any device size in the context of the noise research ($k_F$ is the Fermi wave vector and *n* is the charge carrier concentration). Magnons – quanta of spin waves – typically have wavelength, $\lambda_M$, in the range from tens of nanometers to hundreds of micrometers, and as such, retain their essential wave nature in the magnonic devices[2,5,19]. This fundamental difference is expected to affect the random fluctuation processes leading to noise in magnon currents. Understanding the noise characteristics of magnons, particularly at room temperature, is critical for further development of magnon spintronic technology.

In this study, we focus on the *amplitude* noise of magnons, which sets the limits of the performance of the magnonic devices for information processing or sensing. The experiments are intentionally conducted on an *archetypal* spin waveguide – main element of all magnonic devices, which utilize pure spin wave currents. We do not consider noise in magnetic spin tunneling structures. The schematic of our waveguide structure is shown in Figure 1a. It consists of an electrically insulating yttrium iron garnet (YIG; $Y_3Fe_2(FeO_4)_3$) magnetic waveguide with two micro-strip antennas fabricated directly on top of its surface. One of the antennas is used for magnon excitation by applying RF current. The alternating electric current produces a non-uniform alternating magnetic field around the conducting contour, which, in turn, generates spin waves in the YIG channel under the spin wave resonance conditions. The second antenna is used to measure the inductive voltage produced by the spin wave, *i.e.* magnon current, propagating in the YIG waveguide. The details of the structure and measurements can be found in *METHODS*.

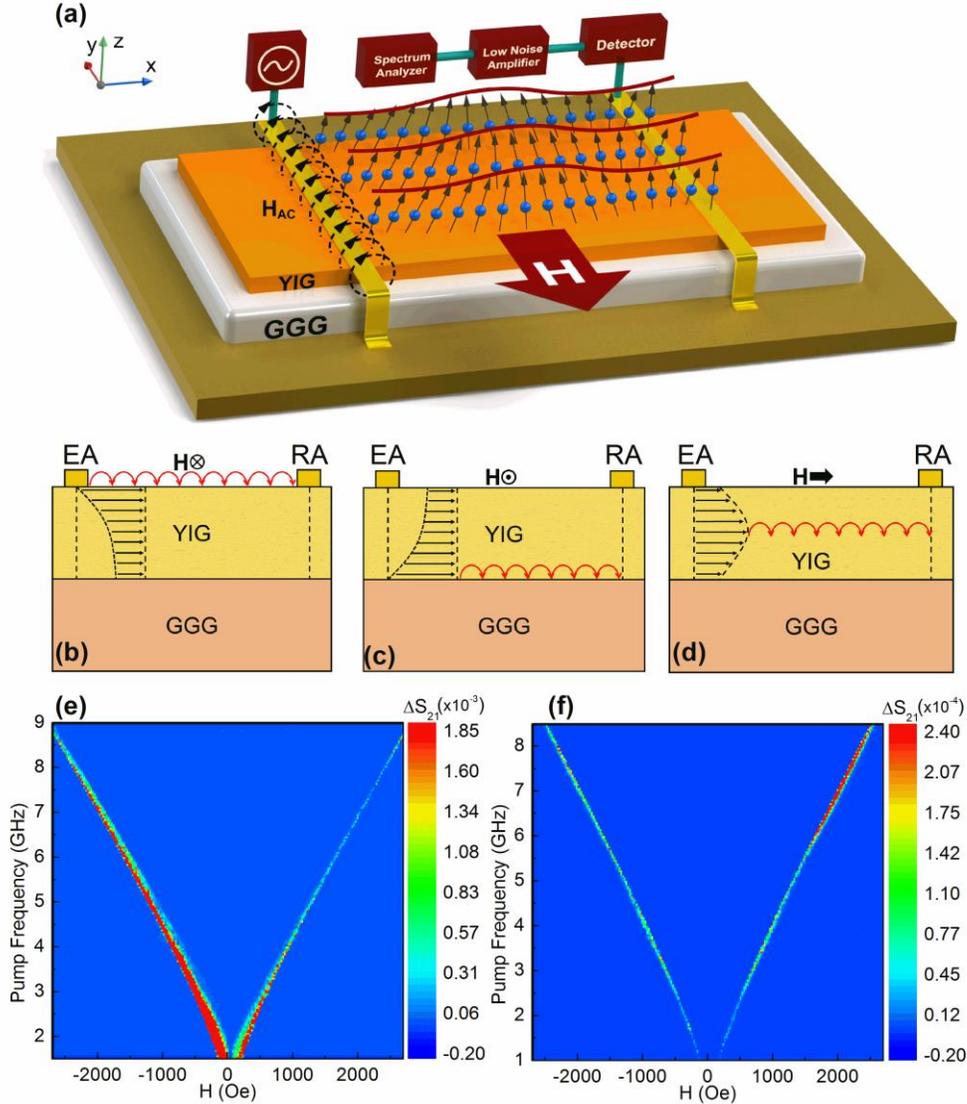

**Figure 1: Spin waveguide structure and types of magnons**. (a) Schematic of the device structure showing the spin waveguide, transmitting and receiving antennas, as well as connection of the noise measurement equipment. (b), (c), (d) Illustration of the propagation of the surface, interface and volume magnon currents, respectively. (e) The normalized scattering parameter ($\Delta S_{21}$) for the surface (left dispersion branch; negative $H$) and interface (right dispersion branch; positive $H$) magnons as a function of the frequency and external DC magnetic field. The dark blue color represents a low output response and the red color a higher output response, corresponding to the propagating magnons. (f) The normalized scattering parameter ($\Delta S_{21}$) for the volume magnons.

We start by confirming the generation and propagation of magnon current through the electrically insulating waveguide. If the bias magnetic field, $H$, is directed in-plane, along the direction of propagation, the spin waveguide structure supports the backward volume magneto-static spin wave (BVMSW)[20]. If $H$ is directed in-plane, orthogonal to the magnon propagation, the structure supports the magneto-static surface spin waves (MSSW). There are two surfaces for MSSW propagation: the top surface of the YIG waveguide and the interface between the YIG waveguide and GGG substrate (see Figures 1b, 1c, 1d). The maximum of the spin wave amplitude is either on the top surface or at the interface depending on the orientation of $H$. Below, we referrer to the three described types of spin waves, and corresponding magnons, as, *surface, interface, and volume*. The wave vector, $k$, of magnon depends on the excitation pump frequency, $f_p$, and $H$. Different types of magnons can propagate through the waveguide only at certain combinations of $k$, $f_p$ and $H$. The magnon current reveals itself in the change of the transmission parameter $S_{21}$[Refs. [3,5,21]]. Figures 1e and 1f show the normalized scattering parameters $\Delta S_{21}$ ($\Delta S_{21} = S_{21}(H) - S_{21}(H = 0)$) for surface and volume spin waves, *i.e.* magnon currents as a function of frequency and magnetic field. The normalization procedure allows us to distinguish the spin wave contribution from other effects[3,9,21]. The measured dispersion data were well fitted with the known dispersion for BVMSW and MSSW, confirming the type of propagating magnons, and allowing for tuning the $f_p - H$ space parameters for the magnon noise studies. To minimize the magnon damping, we selected the pump frequency, $f_p = 5.3\ GHz$, in the frequency range where the three-magnon dissipation processes are prohibited: $f_p > f_{th}^{3m}$ [Ref. 22]. Here $f_{th}^{3m} = \gamma 4\pi M_0 \approx 4.9\ GHz$ is the maximum allowable pump frequency for the three-magnon decay, $\gamma = 2.8\ MHz/Oe$ and $4\pi M_0 = 1750\ G$ is the saturation magnetization of our YIG film.

Propagating in the waveguide, magnon current acquires variations in the amplitude and phase due to the fluctuations of the physical properties of the YIG thin film. To measure these fluctuations, we connected the Schottky diode detector to the receiving antenna (see Figure 1a). The DC signal from the diode was amplified by the low-noise amplifier and recorded with the spectrum analyzer. As a result, the amplitude noise spectrum of magnons was obtained. The noise was studied separately for the different types of magnons, *i.e.* spin waves: surface, interface, and volume at frequency of analysis, $f_a$, ranging from 1 $Hz$ to ~10 $kHz$. Figure 2a shows attenuation as a function of the input power $P_{in}$. In the linear, low-power regime, $P_{in} < 5\ dBm$, the losses were practically independent of the excitation power and the amplitude noise was below the system sensitivity. In this regime, the noise of magnons expressed in the normalized noise spectral density, $S_V/V^2$, was below $10^{-11}\ Hz^{-1}$ ($V$ is the DC voltage on the Schottky diode). Figure 2b presents the noise of the surface, volume and interface magnons at higher input power at $f_a = 10\ Hz$. The first observation is that the noise of the interface magnons was generally higher than that of the surface magnons. The latter was explained by the YIG/GGG interface roughness, resulting in stronger fluctuations of the material parameters that govern magnon current propagation. The noise of the volume magnons, generally was the lowest, and reveled only a moderate increase with $P_{in}$. The increase in $P_{in}$ to some threshold power level, $P_{in} = P_T$, resulted in the abrupt increase of the noise of the interface and surface magnons. One should note that even at their peak values, the noise level of magnons was still relatively low. Similar noise levels of $10^{-9}\ Hz^{-1} - 10^{-5}\ Hz^{-1}$ are found in conventional transistors and other electronic devices[23,24]. However, such devices have orders of magnitude smaller area. In majority of cases, in electronic devices, the low-frequency noise scales inversely proportional to the number of the fluctuators in a device, *i.e.* inversely proportional to its volume or area. Below we show that, surprisingly, this scaling law is not applicable for magnonic devices.

The noise spectra of the surface and interface magnons reveal an intriguing non-monotonic behavior with the input power. Our numerous experiments with several waveguides have shown that $P_T$ of the onset of the high magnon noise corresponds to the point of a strong increase of the attenuation, *i.e.* the offset of the nonlinear dissipation. Given that the pumping frequency $f_p > f_{th}^{3m}$, the dominant nonlinear dissipation mechanism should be related to the four-magnon processes[22,25]. A rough estimate of the $P_{in}$ value when the four-magnon process for MSSW become allowable can be made as[25] $P_{th}^{4m} \approx m_{th}^2 V_g w d$, where $V_g$ is the magnon group velocity, $w$ is the film width, $d$ is the film thickness, $m_{th}$ is the threshold amplitude of variable magnetization defined by the film magnetic properties. Using measured $V_g = 1.25 \cdot 10^7\ cm/s$, geometry and $m_{th}$ from literature[22,25], we estimate $P_{th}^{4m} \approx 0.24\ mW$, which is below the typical pumping level in our experiments. The latter indicates that for the selected $f_p$ and $P_{in} \geq P_{th}^{4m}$, the four-magnon processes are allowed in our system.

In the four-magnon scattering, two magnons of frequency $f_p$ annihilate and create a pair of quickly dissipating magnons of close frequencies and counter directed $k$-vectors[11,22]. At some density of initial magnons these processes become avalanche-like, leading to a sharp decay of the initial magnon current. The first peaks in the noise spectral density of the surface and interface magnon currents appears at the onset of the avalanche-like four-magnon processes (~8 $dBm$ for surface and ~10 $dBm$ for interface) when the system is fluctuating between the linear and non-linear regimes (Figure 2 (b)). Further increase in $P_{in}$ leads to reduction in the noise when the system stabilizes to a certain type of four-magnon processes followed by the second peaks, which likely correspond to on-set of other types of four-magnon processes, *e.g.* with different wave-vectors and frequencies[22]. It is interesting to note that the noise signatures of the four-magnon processes – abrupt few-orders-of-magnitude peaks (Figure 2 (b)) – are much more clear than the amplitude attenuation signatures – corresponding gradual changes in the slopes (Figure 2 (a)). Until now, most of the nonlinear magnon scattering was studied using the time-resolved magneto-optical Kerr effect[26] or Brillouin spectroscopy[27]. The demonstrated ability to monitor nonlinear magnon damping phenomena via noise spectroscopy provides a powerful tool for studying multi-magnon processes.

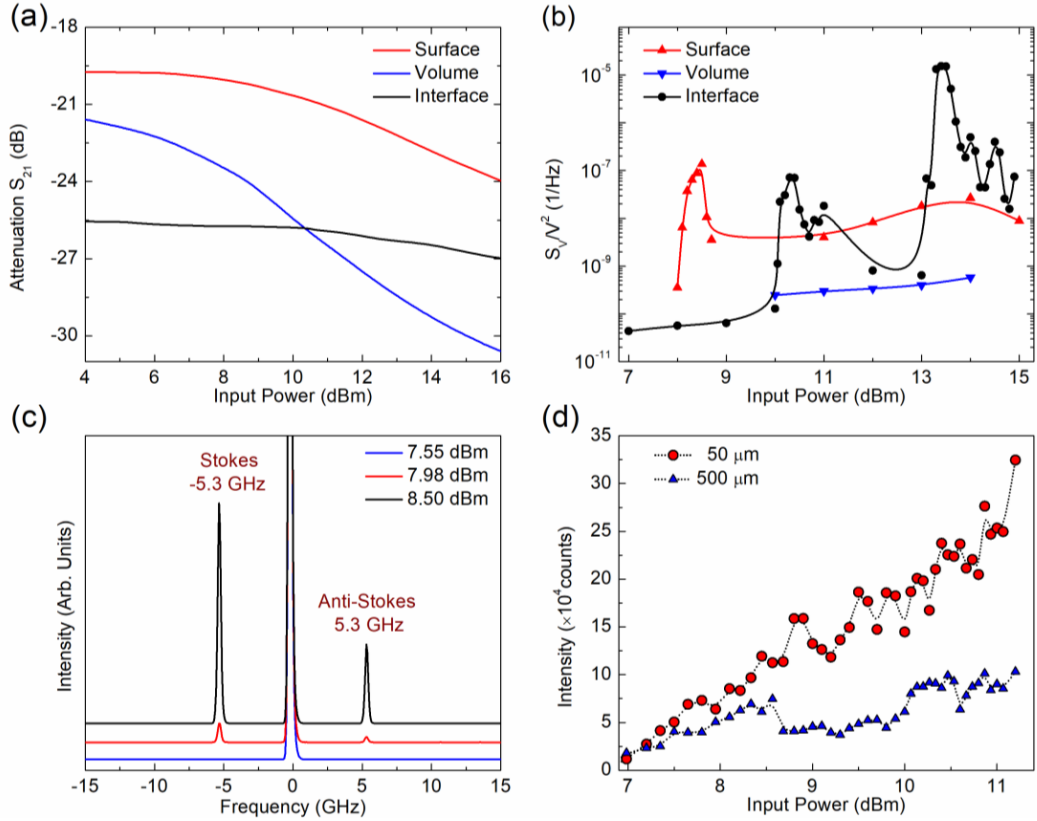

**Figure 2: Propagation and noise of magnons.** (a) Attenuation of the surface, volume and interface magnons as a function of the excitation power at the excitation frequency of $f_p \approx 5.3\ GHz$. The normalized noise spectral density, $S_V/V^2$, of the voltage fluctuations at the frequency of the analysis of $f_a = 10\ Hz$. Note an abrupt increase in the noise of surface and interface magnons as the input power reaches some threshold level. The noise of volume magnons does not change substantially. (c) Brillouin light scattering spectrum of surface magnons. The magnon peaks as seen at the excitation frequency set by the generator. (d) The intensity of magnon Stoke peak as the function of the input power for two locations measured from the position of the transmitting antenna. The on-set of strong intensity fluctuations corresponds to the abrupt increase in noise level and on-set of nonlinear dissipation associated with the four-magnon processes.

To provide independent confirmation of the correlation of the magnon current noise with the onset of strong multi-magnon processes, we conducted *in-situ* Brillouin − Mandelstam spectroscopy (BMS) of the propagating magnons[28,29]. In these experiments, we focused laser light on YIG channel and varied the input RF power. The details of the measurements can be found in *METHODS*. Figure 2c shows the representative BMS spectra with clear Stoke and anti-Stoke signatures of the surface magnons at the pump frequency. The intensity of the magnon peaks increases strongly with the increasing $P_{in}$. The propagation of magnons and the on-set of nonlinear dissipation can be observed from the plot of the magnon intensity as the function of input power at two locations along the waveguide (Figure 2d). At certain power level the dependence becomes strongly non-monotonic and unstable, indicating the on-set of nonlinear dissipation, in line with the noise spectroscopy data (Figure 2b).

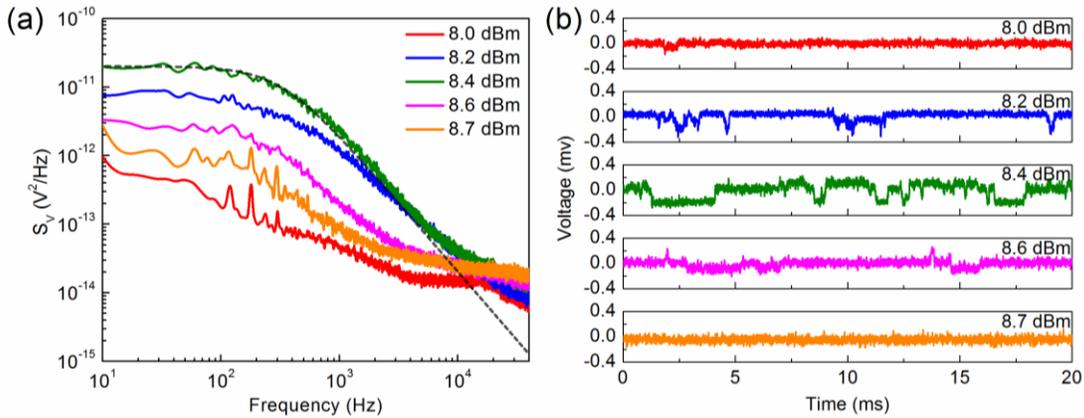

**Figure 3: Magnon noise and discrete fluctuators.** (a) Noise spectra for five slightly different input power levels, corresponding to the first noise maxima for surface magnons in Figure 1b. The low-frequency noise has pronounced Lorentzian characteristics, which is in striking contrast to $1/f$ noise in macroscopic electronic devices. (b) The random telegraph signal noise of magnons shown in the time domain for the same input power levels. The data are presented for $f_p \approx 5.3\ GHz$. The well-defined RTS noise is an indication that a single discrete fluctuator makes a dominant contribution to noise.

The magnon noise spectrum as a function of frequency revealed another unusual feature. The spectra of the amplitude fluctuations had the shape of the clear Lorentzian, $S_V \sim 1/(1 + f^2/f_c^2)$ with the characteristic corner frequency $f_c < 100 - 1000\ Hz$ (see Figure 3a). This is in striking contrast to macroscopic electronic devices where the low-frequency noise is usually dominated by either $1/f$ noise or its superposition with G-R noise[17,18,30]. In the time domain, the magnon noise revealed itself as a random telegraph signal (RTS) noise, appearing as series of pulses of the fixed amplitude and random pulse width and time intervals between the pulses. Representative recordings of RTS noise of magnons are shown in Figure 3 (b). Interestingly, very small changes in $P_{in}$ resulted in significant changes in the magnon noise characteristics. The level and shape of the noise spectra as well as the shape of the RTS traces changed strongly with the power input. The RTS noise is well known in semiconductor devices[30]. It appears when a single fluctuator makes a dominant contribution to noise. For example, in a field-effect transistor with a very small gate area, RTS is due to the capture and emission of an electron by a *single* trap[31]. Our observation of RTS noise in the large magnon waveguides suggests that in the nonlinear dissipation regime, individual *discrete* macro events contribute to both the noise and magnon dissipation processes. We found this kind of RTS noise in *all* studied magnonic devices, which suggests that this is a specific feature of nonlinear dissipation process of magnons. At high frequencies, the noise decreases as $1/f^2$ and falls below the background noise level.

The dominance of RTS noise in magnon devices has important implications for the device scaling. We established that a single discrete fluctuator contributes to the noise peaks. The reduction of the device area does not change the number of fluctuators until the device area is of the order of the dimension of the fluctuator itself. The size of the fluctuator can be roughly estimated if we assume that this fluctuator blocks the spin wave completely. Following the analogy with the charge density waves[32], we can write that $\delta A/A = \delta V/V$, where $\delta A$ is the area of the fluctuator, $A = w \times L_D$ is the total active area of the device, $w$ is the width of the YIG waveguide, and $L_D$ is the characteristic length, related to the wave attenuation, and $\delta V/V \approx 10^{-2}$ is the relative amplitude of the RTS noise. This estimate yields the area $\approx 10^4 \ \mu m^2$. In comparison with semiconductor devices, this is a very large area. However, it is still orders of magnitude smaller than the total area of the studied YIG waveguide. The latter means that the $A^{-1}$ scaling does not apply for magnons and one can fabricate much smaller magnonic devices without automatic increase in the magnon noise. The wavelength of spin waves in our experiments can be estimated as $\lambda_M = 2\pi L/\Psi_T$, where $\Psi_T$ is the total phase difference accumulated over the propagation distance between antennas, $L$. With $\Psi_T$ directly measured by VNA , we obtained $\lambda_M \approx 390 \ \mu m$, which confirms the large spatial extend of magnons in our experiments.

The concentration of magnons in our devices, $n_M$, can be estimated *as* $n_M = m^2/(2M_0\gamma h)$, where $m$ is the variable magnetization, $M_0$ is the saturation magnetization, $\gamma = 2.8 \ MHz/Oe$, and $h$ is Plank's constant[22]. Taking $M_0 = 139 \ Oe$ for YIG, we obtain $n_M \approx 2 \times 10^{17} \ cm^{-3}$. For a given magnon concentration, the large amplitude of the RTS signal, $\delta V/V \approx 10^{-2}$, indicates that a large number of magnons disappear during a single step of the RTS signal. The latter suggests an extremely unusual RTS noise of magnons as compared to electrons in electronic devices. For example, in downscaled metal-oxide-semiconductor field-effect transistors, a single RTS step can correspond to just *one* electron captured by the trap[17,30,31]. Therefore, the RTS-like signal found in magnonic devices is different from classical RTS noise in semiconductor devices and has certain inherent discreteness revealed via individual fluctuation events of spatially large fluctuators with unusually high amplitude. Our data suggests that the mechanism of this *discrete* noise corresponds to the four-magnon dissipation processes. When the excitation power is close to some threshold value for these processes, small fluctuations in the YIG physical parameters lead to the avalanche-like multi-magnon processes involving an exceptionally high number of magnons.

In conclusion, we investigated the noise of magnon currents in electrically insulating spin waveguides. It was discovered that the low-frequency amplitude noise of magnons is dominated by RTS noise, unlike the noise of electrons in conventional devices, which is mostly dominated by $1/f$ noise. Our findings suggest that the noise of wave-like magnons, characterized by a large spatial extend and exceptionally large number of magnons participating in each RTS step, reveals an unusual *discrete* nature. It is also rather muted, or *discreet*, at the lower powers levels. We have established that the volume magnons produce much less noise than surface and interface magnons. The noise of surface and interface magnons increase sharply at the on-set of nonlinear avalanche-like four-magnon processes. It was also demonstrated that noise spectroscopy can serve as a valuable tool for investigating non-linear magnon dissipation.

**METHODS**

**Device structure and experimental procedures:** The test structure consisted of a $Y_3Fe_2(FeO_4)_3$ thin-film spin waveguide epitaxially grown on top of a gadolinium gallium garnet ($Gd_3Ga_5O_{12}$) substrate by the liquid phase epitaxy. The thickness of the YIG film was $d = 9.6\,\mu m$, and its saturation magnetization was $4\pi M_0 \approx 1750\,Oe$. Two micro-strip antennas with the width of 100 μm and thickness of 100 nm were fabricated directly on top of the YIG surface near the waveguide edges. The antennas were orientated perpendicular to the axis of waveguide. The center-to-center distance between the antennas was $L = 7.5\ mm$, the width of the waveguide channel was $w = 2\ mm$. The device under test was placed inside an electromagnet (GMW 3472) with the pole cap 50-mm diameter tapered to provide a uniform bias magnetic field with $\Delta H/H < 10^{-4}$ per 1 mm in the range from $-2500\ Oe$ to $+2500\ Oe$. The antennas were connected to a network analyzer (Keysight PNA N5221A). One of the antennas was used for spin wave excitation by applying RF current. The second antenna was used to measure the inductive voltage produced by the spin wave propagating in the YIG waveguide. The details of the inductive measurement technique can be found elsewhere[33]. The dispersion of MSSW (see Figure 2b) can be written as[34]:

$$f_{MSSW} = \gamma\sqrt{H(H + 4\pi M_0) + (4\pi M_0)^2(1 - \exp(-2kd)/4)} \tag{M1}$$

Here $\gamma = 2.8\ MHz/Oe$ is the gyromagnetic ratio, $H$ is the bias in-plane magnetic field, $k$ is the in-plane wave vector, and $d$ is the thickness of the film. The experimental data were well-fitted with Eq. (M1). The non-reciprocal nature of MSSW propagation reveals itself in the difference between the scattering parameters $S_{21}$ and $S_{12}$. The surface mode has a non-reciprocal behavior in the spin wave amplitude for the waves with opposite signs of the wave vectors ($\pm k$) [Refs. 34–37]. The waves propagate on the opposite (top surface *vs.* bottom interface) sides of the YIG waveguide (see Figure 1). The dispersion of BVMSW (see Figure 2c) can be written as[34]:

$$f_{BVMSW} = \gamma\sqrt{\left(H^2 + H4\pi M_0(1 - \exp(-kd)/kd)\right)} \tag{M2}$$

The experimental data in Figure 1c were, again, well-fitted with Eq. (M2), assuming $k$ in the range from 0 to 314 cm$^{-1}$. The experiments and numerical simulations confirmed the type and nature of the spin wave, *i.e.* magnon current. More details on the magnon waveguide propagation and measurements procedures can be found in *SUPPLEMENTAL INFORMATION*.

**Noise measurements:** The YIG-film waveguide with antennas was placed in a brass sample holder with SMA connectors for antennas. Antennas were shortened to the sample holder at one end and connected to the SMA connector at another end. In order to measure the S parameters, the antennas were connected to the vector network analyzer (VNA) (Keysight Technologies PNA N5221A). For the noise measurements, the spin waves were excited by the excitation antenna powered by the VNA in the continues wave (CW) operation mode. Propagation along the waveguide the spin waves acquires fluctuations in the amplitude and phase due to fluctuations in the physical parameters' values inside the YIG crystal. In order to measure the amplitude fluctuations, a Schottky diode (Keysight Technologies 33330B) was connected to the receiver antenna. The DC detected signal from the diode was amplified by the low noise amplifier (Stanford Research 560) and analyzed by a spectrum analyzer (Bruel & Kjaer FFT Photon). As a result, the spectrum of the low-frequency amplitude fluctuations at given RF frequency and power of the excitation was obtained. We excluded the background electronic noise of the detector and RF generator by conducting controlled experiments with the detector connected directly to the VNA output. In all experiments, the level of the background noise was at least an order of magnitude smaller than that measured noise of magnons.

**Brillouin – Mandelstam spectroscopy of spin waves:** BMS is an optical technique, which can be used to detect phonons and magnons in the frequency range from 1 GHz to 900 GHz, and the wave vectors close to the Brillouin zone center[27–29,38]. The experiments were carried out in backscattering configuration at the normal incident of the laser light with respect to the surface of the sample. The light source was a continuous-wave solid-state diode-pumped laser (Coherent, V-2) operating at $\lambda = 532\ nm$. The light was focused on the sample using a lens with the *f*-number of $N = 1.4$. The scattered light was collected with the same lens and directed to the high-resolution high-contrast six-pass tandem Fabry-Perot interferometer (TFP-1; JRS instruments). The polarization of the incident and scattered light was analyzed carefully. The incident laser light is *p*-polarized with high polarization purity (extinction ratio of 100000:1). Since the spin waves rotate the polarization of the incident light by 90 degrees as they scatter the light, only the scattered light with linearly *s*-polarization was directed to the interferometer using a high extinction ratio (100000:1) polarizer (Glan-Laser Calcite polarizer; Thorlabs Inc.). During the experiments, the position of the laser spot was fixed. The position of the spot was carefully monitored in order to avoid any displacement due to the temperature drift. In order to avoid self-heating effects, the power of the incident light was adjusted to $\sim 1\ mW$. Owing to the high signal-to-noise ratio of the spectrum, short accumulation time (~3 minutes) was sufficient to obtain accurate data. More details of the measurements can be found in *SUPPLEMENTAL INFORMATION*.

**Acknowledgements**

The work at UC Riverside was supported as part of the Spins and Heat in Nanoscale Electronic Systems (SHINES), an Energy Frontier Research Center funded by the U.S. Department of Energy, Office of Science, Basic Energy Sciences (BES) under Award # SC0012670. The authors are indebted to H. Chiang for performing metal deposition, J.S. Lewis and Z. Barani for assistance with BMS measurements, E. Hernandez for his help with the device schematics.

**Contributions**

A.A.B. conceived the idea, coordinated the project and contributed to the data analysis; S.R. conducted noise measurements and data analysis; M.B. fabricated the devices and contributed to noise measurements; A.K. supervised device fabrication and contributed to the data analysis; F.K conducted BMS measurements. A.A.B. led the manuscript preparation. All authors contributed to writing and editing of the manuscript.